\documentstyle[amsfonts,12pt,thmsa,a4,sw20aip]{article}

\input{tcilatex}
\begin{document}

\title{Remarks on Lewenstein-Sanpera Decomposition}
\author{{\small Mingjun Shi\thanks{%
Email: shmj@ustc.edu.cn}, Jiangfeng Du} \\
{\small Laboratory of Quantum Communication and Quantum Computation,}\\
{\small \ Department of Modern Physics,}\\
{\small \ University of Science and Technology of China, Hefei, 230026, 
P.R.China}}
\date{Dec. 27, 2000}
\maketitle

\begin{abstract}
We discuss in this letter Lewenstein-Sanpera (L-S) decomposition for a
specific Werner state. Compared with the optimal case, we propose a
quasi-optimal one which in the view of concurrence leads to the same
entanglement measure for the entangled mixed state discussed. We think that
in order to obtain entanglement of given state the optimal L-S decomposition
is not necessary.
\end{abstract}

Non-locality, entanglement or inseparability are some of the most genuine
quantum concepts\cite{1,2,3}, and play a very important role in quantum
computation and quantum communication\cite{4,5,6}. To study the entanglement
phenomena embodied in mixed states is an intricate work. Associated with
different definitions of the entanglement of mixed states, various
quantitative measures have been proposed\cite{6,7,8,9,10}. Among them is
Lewenstein-Sanpera (L-S) decomposition\cite{9}. It is said in Ref\cite{9}
that any density matrix $\rho $ in ${\Bbb C}^2\times {\Bbb C}^2$ can be
decomposed as 
\begin{equation}
\rho =\lambda \rho _s+\left( 1-\lambda \right) P_e,\quad \lambda \in \left[
0,1\right] ,  \label{1}
\end{equation}
where $\rho _s$ is separable density matrix, $P_e$ denotes a single pure
entangled projector $\left( P_e\equiv \left| \Psi _e\right\rangle
\left\langle \Psi _e\right| \right) $. Given $\rho $ there are many
different $\rho _s$'s and $P_e$'s satisfying Eqn.(1). The optimal case is
unique in which $\lambda $ is maximal, that is 
\[
\rho =\lambda ^{(opt)}\rho _s^{(opt)}+(1-\lambda ^{(opt)})P_e^{(opt)}.
\]
Any other decomposition of the form $\rho =\widetilde{\lambda }\widetilde{%
\rho _s}+\left( 1-\widetilde{\lambda }\right) \widetilde{P_e},$ with $%
\widetilde{\lambda }\in \left[ 0,1\right] $ such that $\widetilde{\rho _s}%
\neq \rho _s^{(opt)}$ necessarily implies that $\widetilde{\lambda }<\lambda
^{(opt)}$. Then due to the uniqueness, the decomposition given by Eqn.(1)
leads to an unambiguous measure of the entanglement for any mixed state $%
\rho $, that is 
\begin{equation}
E(\rho )=(1-\lambda ^{(opt)})\times E(\left| \Psi _e\right\rangle ^{(opt)}),
\end{equation}
where $E(\left| \Psi _e\right\rangle ^{(opt)})$ is the entanglement of its
pure state expressed in terms of the von Neumann entropy of reduced density
matrix of either of its subsystems. And moreover, L-S point out that this
measure of entanglement is independent of any purification or formation
procedure.

In this letter, we discuss L-S decomposition for a specific entangled
Werner's state. The optimal decomposition is discussed in Ref\cite{9} and Ref%
\cite{11}, where $\lambda $ in Eqn.(1) reaches the maximal $\lambda ^{(opt)}$
while $\left| \Psi _e\right\rangle ^{(opt)}$ denotes the maximal entangled
state, Bell state. But we will be interested in the quasi-optimal
decomposition in which $\left| \Psi _e\right\rangle $ is not Bell state and $%
\lambda $ is maximal relative to the $\left| \Psi _e\right\rangle $. We find
that on the basis of entanglement concurrence\cite{8} the optimal
decomposition and the quasi-optimal one give the same result. The detail is
given as follows.

We have known that a Werner's state can be expressed as 
\begin{equation}
\rho _w=\left( 1-\epsilon \right) \rho _0+\epsilon P_{Bell},\quad \epsilon
\in \left[ 0,1\right] ,
\end{equation}
where $\rho _0$ is the maximal separable state, i.e., $\rho _0=\frac
14I_{4\times 4}$, $P_{Bell}=\left| \Psi _{Bell}\right\rangle \left\langle
\Psi _{Bell}\right| $, and $\left| \Psi _{Bell}\right\rangle $ is one of
four Bell states. If and only if $\frac 13<\epsilon \leq 1$, $\rho _w$ is
inseparable. In this letter, we focus on a specific case in which $\epsilon
=\frac 12$ and $\left| \Psi _{Bell}\right\rangle =\left| \Psi
^{-}\right\rangle =\frac 1{\sqrt{2}}\left( \left| 01\right\rangle -\left|
10\right\rangle \right) $. That is, 
\begin{equation}
\rho _{\frac 12}=\frac 12\rho _0+\frac 12\left| \Psi ^{-}\right\rangle
\left\langle \Psi ^{-}\right| =\frac 14\left( I\otimes I-\frac 12\sigma
_1\otimes \sigma _1-\frac 12\sigma _2\otimes \sigma _2-\frac 12\sigma
_3\otimes \sigma _3\right) ,
\end{equation}
where $I$ is $2\times 2$ identity matrix, and $\sigma _i$ is pauli matrix.

The optimal L-S decomposition has been obtained in Ref\cite{11}, that is,
for $\rho _{\frac 12}$ expressed by (4), $\lambda ^{(opt)}=\frac 34$, $%
P_e^{(opt)}=\left| \Psi ^{-}\right\rangle \left\langle \Psi ^{-}\right| $,
and $\rho _{\frac 12}$ can be written as 
\begin{equation}
\rho _{\frac 12}=\frac 34\rho _s^{(opt)}+\frac 14\left| \Psi
^{-}\right\rangle \left\langle \Psi ^{-}\right| ,
\end{equation}
where $\rho _s^{(opt)}=\frac 23\rho _0+\frac 13\left| \Psi ^{-}\right\rangle
\left\langle \Psi ^{-}\right| $, and evidently $\rho _s^{(opt)}$ is
separable. According to Eqn.(2), the entanglement of $\rho _{\frac 12}$ is 
\begin{equation}
E\left( \rho _{\frac 12}\right) =\frac 14E(\left| \Psi ^{-}\right\rangle
)=\frac 14.
\end{equation}

Now we consider a decomposition of a more general form, 
\begin{equation}
\rho _{\frac 12}=\delta \rho _s+\left( 1-\delta \right) \left| \Psi
\right\rangle \left\langle \Psi \right| ,\quad \delta \in \left[ 0,1\right] ,
\end{equation}
where we assume $\left| \Psi \right\rangle =\cos \theta \left|
01\right\rangle -\sin \theta \left\langle 10\right| ,\ \theta \in \left[
0,\frac \pi 2\right] $. We wish to find the maximal $\delta $ relative to $%
\left| \Psi \right\rangle $. This decomposition can be called quasi-optimal
decomposition.

Let $x=\frac 1\delta -1$. $x$ is non-negative. We rewrite Eqn.(7) as 
\begin{equation}
\rho _s=\left( 1+x\right) \rho _{\frac 12}-x\left| \Psi \right\rangle
\left\langle \Psi \right| ,\quad x>0.
\end{equation}
In other words, $\rho _s$ is the pseudo-mixture of $\rho _{\frac 12}$ and $%
\left| \Psi \right\rangle \left\langle \Psi \right| $. Furthermore, we give
the detailed form of Eqn.(8). 
\begin{eqnarray}
\rho _s &=&\frac 14[I\otimes I-x\cos 2\theta \ \sigma _3\otimes I+x\cos
2\theta \ I\otimes \sigma _3  \nonumber \\
&&+\left( x\sin 2\theta -\frac 12\left( 1+x\right) \right) \sigma _1\otimes
\sigma _1  \nonumber \\
&&+\left( x\sin 2\theta -\frac 12\left( 1+x\right) \right) \sigma _2\otimes
\sigma _2  \nonumber \\
&&-\frac 12\left( 1-x\right) \sigma _3\otimes \sigma _3].
\end{eqnarray}
Then the problem is to find the minimal $x$ such that $\rho _s$ is a
separable state. To guarantee the positivity of $\rho _s$, we should know
the eigenvalues of $\rho _s$. Or we can use the positivity criteria in Ref%
\cite{11}. To determine the separability of $\rho _s$, we shall consider the
positivity of the partial transposition of $\rho _s$ ($\rho _s^{T_A}$ or $%
\rho _s^{Y_B}$) or the partial time-reversal of $\rho _s$ ($\widetilde{\rho
_s}$)\cite{12,13,14,15}. Again the criteria in Ref\cite{11} can be used to
verify the positivity of $\widetilde{\rho _s}$. Through laborious but not
difficult mathematical computation we have the following results:

(i) Decomposition of the form (7) can be accomplished only if $\sin 2\theta
\geq \frac 7{12}$. When $\sin 2\theta <\frac 7{12}$, $\rho _s$ expressed by
(8) is either non-positive or inseparable.

(ii) Under the condition $\sin 2\theta \geq \frac 7{12}$ satisfied, the
minimal $x$ which ensures positivity and separability of $\rho _s$ in
Eqn.(9) is 
\begin{equation}
x_{min}=\frac 1{4\sin 2\theta -1}.
\end{equation}
Correspondingly, the maximal $\delta $ is 
\begin{equation}
\delta _{max}=1-\frac 1{4\sin 2\theta }.
\end{equation}
The $\delta _{max}$ is maximal relative to a proper entangled state $\left|
\Psi \right\rangle $ and is just what we want to know to realize the
quasi-optimal decomposition of $\rho _{\frac 12}$ in the form Eqn.(7).

\medskip

\bigskip

Now let's discuss our results. First, in general sense, an arbitrary
entangled state can not necessarily be used as the component of L-S
decomposition of Eqn.(1). From the specific example discussed in this
letter, we see that the entanglement of the pure state appearing in the
decomposition can not be too small. Then, recall that the concept of
concurrence for pure state\cite{8}. For any pure state in ${\Bbb C}^2\times 
{\Bbb C}^2$ described as 
\begin{equation}
\left| \psi \right\rangle =c_0\left| 00\right\rangle +c_1\left|
01\right\rangle +c_2\left| 10\right\rangle +c_3\left| 11\right\rangle ,
\end{equation}
where $c_i$'s are complex numbers and satisfies $\sum_{i=0}^3\left|
c_i\right| ^2=1$, the concurrence is defined as 
\begin{equation}
C\left( \left| \psi \right\rangle \right) =2\left| c_0c_3-c_1c_2\right| .
\end{equation}
And the entanglement can be expressed in terms of $C\left( \left| \psi
\right\rangle \right) $, i.e., 
\begin{equation}
E\left( \left| \psi \right\rangle \right) =-\frac{1+\sqrt{1-C^2}}2\log _2%
\frac{1+\sqrt{1-C^2}}2-\frac{1-\sqrt{1-C^2}}2\log _2\frac{1-\sqrt{1-C^2}}2
\end{equation}
The concurrence of Bell state $\left| \Psi ^{-}\right\rangle $ is $C\left(
\left| \Psi ^{-}\right\rangle \right) =1$, and that of $\left| \Psi
\right\rangle $ in Eqn.(7) is $C\left( \left| \Psi \right\rangle \right)
=2\sin \theta \cos \theta =\sin 2\theta $. So for the optimal L-S
decomposition of $\rho _{\frac 12}$, we have 
\begin{equation}
\left( 1-\lambda ^{\left( opt\right) }\right) C\left( \left| \Psi
^{-}\right\rangle \right) =\frac 14.
\end{equation}
For quasi-optimal decomposition of $\rho _{\frac 12}$ with the form (7), we
have 
\begin{equation}
\left( 1-\delta _{max}\right) C\left( \left| \Psi \right\rangle \right)
=\left[ 1-(1-\frac 1{4\sin 2\theta })\right] \sin 2\theta =\frac 14.
\end{equation}
The same result of (15) and (16) means that at least for this specific $\rho
_{\frac 12}$, the optimal and quasi-optimal decomposition can be used to
demonstrate the entanglement proportion embodied in $\rho _{\frac 12}$ in
terms of concurrence. Because of the uniqueness, the optimal L-S
decomposition indicates more strict constraints, and it is a hard work to
find it. Comparatively speaking, the quasi-optimal decomposition is easy to
realize and may be a convenient method to study entangled mixed states. On
the other hand, we easily see that for $\rho _{\frac 12}$%
\begin{equation}
\left( 1-\lambda ^{\left( opt\right) }\right) E\left( \left| \Psi
^{-}\right\rangle \right) =\frac 14\neq \left( 1-\delta _{max}\right)
E\left( \left| \Psi \right\rangle \right) .
\end{equation}
That is, in term of von Neumann entropy, there is inconsistence between the
optimal and the quasi-optimal. Note Eqn.(14). $E\left( \left| \psi
\right\rangle \right) $ is a logarithmic function of $C$. We consider that
logarithm conceals the agreement on the level of concurrence. So we think
that concurrence is the proper quantity to measure the entanglement of mixed
states in the frame of L-S decomposition. In our view, obtaining the
quasi-optimal decomposition is sufficient to measure the entanglement
proportion in a mixed state.

Of course, in this letter we have only studied a specific example. We wish
to extend our discussion to more general cases. Further results will be
submitted later.

\bigskip

\bigskip {\bf ACKNOWLEDGMENTS}

This project is supported by the National Nature Science Foundation of China
(10075041 and 10075044) and the Science Foundation of USTC for Young
Scientists.


\bigskip

\begin{thebibliography}{99}
\bibitem{1}  A. Einstein, B. Podolsky and N. Rosen, Phys. Rev. {\bf 47}, 777
(1935).

\bibitem{2}  E. Schr\"{o}dinger, Proc. Cambridge Philos. Soc. {\bf 31}, 555
(1935).

\bibitem{3}  J. S. Bell, Physics {\bf 1}, 195 (1964).

\bibitem{4}  C. H. Bennett and S. J. Wiesner, Phys. Rev. Lett. {\bf 69},
2881 (1992).

\bibitem{5}  C. H. Bennett, G. Brassard, C. Cr\'{e}peau, R. Jozsa, A. Peres,
and W. K. Wootters, Phys. Rev. Lett. 70, 1895 (1993).

\bibitem{6}  C. H. Bennett, D. P. DiVincenzo, J. A. Smolin, and W. K.
Wootters, Phys. Rev. A54, 3824 (1996).

\bibitem{7}  V. Vedral, M. B. Plenio, M. A. Rippin and P. L. Knight, Phys.
Rev. Lett. {\bf 78}, 2275 (1997).

\bibitem{8}  W. K. Wootters, Phys. Rev. Lett. {\bf 80}, 2245 (1998).

\bibitem{9}  M. Lewenstein and A. Sanpera, Phys. Rev. Lett. {\bf 80}, 2261
(1998).

\bibitem{10}  G. Vidal and R. Tarrach, Phys. Rev. {\bf A59}, 141 (1999).

\bibitem{11}  B. Englert and N. Metwally, e-print archive: quant-ph/9912089.

\bibitem{12}  A. Peres, Phys. Rev. Lett. {\bf 77}, 1413 (1996).

\bibitem{13}  M. Horodecki, P. Horodecki, R. Horodecki, Phys. Lett. {\bf A223%
}, 1 (1996).

\bibitem{14}  P. Horodecki, Phys. Lett. {\bf A232}, 333 (1997).

\bibitem{15}  A. Sanpera, R. Tarrach and G. Vidal, e-print archive:
quant-ph/9704041.
\end{thebibliography}
\end{document}